\documentclass[aps,twocolumn,showpacs,superscriptaddress,floatfix]{revtex4}

\usepackage{dcolumn}
\usepackage{amsfonts,amsmath,amsthm,amssymb,bm,mathrsfs,bbm,amscd}
\usepackage{graphicx}
\usepackage{setspace}
\usepackage{times}
\usepackage[none]{hyphenat}
\usepackage{color}
\usepackage{ulem}
\usepackage{epsfig}
\usepackage{xcolor}

\begin{document}

\title{Recovery rate affects the effective epidemic threshold with synchronous updating}
\author{Panpan Shu}
\affiliation{Web Sciences Center, University of Electronic
Science and Technology of China, Chengdu 610054, China}
\affiliation{School of Sciences, Xi'an University of Technology, Xi'an 710054, China}

\author{Wei Wang}
\email{wwzqbx@hotmail.com}
\affiliation{Web Sciences Center, University of Electronic
Science and Technology of China, Chengdu 610054, China}
\affiliation{Big data research center, University of Electronic
Science and Technology of China, Chengdu 610054, China}

\author{Ming Tang}
\email{tangminghan007@gmail.com}
\affiliation{Web Sciences Center, University of Electronic
Science and Technology of China, Chengdu 610054, China}
\affiliation{Big data research center, University of Electronic
Science and Technology of China, Chengdu 610054, China}

\author{Pengcheng Zhao}
\affiliation{School of Physics and Optoelectronic Engineering,
Xidian University, Xi'an 710071, China}

\author{Yi-Cheng Zhang}
\affiliation{Department of Physics, University of Fribourg, Chemin du Mus\'{e}e 3, 1700 Fribourg, Switzerland}

\date{\today}

\begin{abstract}
Accurate identification of effective epidemic threshold is essential for understanding epidemic dynamics on complex networks. The existing studies on the effective epidemic threshold of the susceptible-infected-removed (SIR) model generally assume that all infected nodes immediately recover after the infection process, which more or less does not conform to the realistic situation of disease. In this paper, we systematically study the effect of arbitrary recovery rate on the SIR spreading dynamics on complex networks. We derive the theoretical effective epidemic threshold and final outbreak size based on the edge-based compartmental theory. To validate the proposed theoretical predictions, extensive numerical experiments are implemented by using asynchronous and synchronous updating methods. When asynchronous updating method is used in simulations, recovery rate does not affect the final state of spreading dynamics. But with synchronous updating, we find that the effective epidemic threshold decreases with recovery rate, and final outbreak size increases with recovery rate. A good agreement between the theoretical predictions and numerical results are observed on both synthetic and real-world networks. Our results extend the existing theoretical studies, and help us to understand the phase transition with arbitrary recovery rate.
\end{abstract}



\maketitle

{\bf How to accurately predict the effective epidemic threshold has attracted increasing attentions. The existing studies on the epidemic threshold generally suppose the recovery process with a constant recovery rate of 1, while the investigation on the effect of recovery rate is still insufficient. Considering the difference of recovery rate between different real diseases and the accompanying effects on the human health, it is very necessary to predict the effective epidemic thresholds with different recovery rates. In this work, the effect of recovery rate on the effective threshold of epidemic outbreak is systematically studied. We first develop a novel theoretical framework based on the edge-based compartmental theory. The developed theory predicts that recovery rate does not affect the spreading dynamics with asynchronous updating, but with synchronous updating, the effective epidemic threshold decreases with the recovery rate, and the final outbreak sizes increases with the recovery rate for a given effective transmission rate. It should be noted that the SIR epidemic of synchronous updating breaks more easily than asynchronous updating. To verify the accuracy of the theoretical predictions, we numerically predict the effective epidemic threshold using the variability measure on random regular networks, where the numerical results agrees well with the theoretical predictions. Moreover, we investigate how the recovery rate affects the epidemic outbreaks with synchronous updating on scale-free networks and real-world networks, and find the same variation trend of effective epidemic threshold. This work provides us a deep understanding of effective epidemic threshold and would promote further studies on phase transition of epidemic dynamics.}

\section{Introduction}\label{Introduction}
Susceptible-infected-recovered (SIR) model on complex networks have been used to model a wide variety of real epidemic spreading~\cite{Barrat:2008,Albert:2012,Gross}. Examples include the spreads of mumps, varicella, rabies and aids~\cite{Anderson:1992}. In the SIR model, an infected node can transmit a disease to each of its susceptible neighbors with probability $\beta$. At the same time, the infected nodes recover with probability $\mu$. In this context, a critical value of the effective transmission rate $\lambda=\beta/\mu$ (or the effective epidemic threshold $\lambda_c$) exists above which the final fraction of recovered nodes is finite~\cite{Pastor-Satorras:2014,Boccaletti2006}.

In previous studies, it is pointed out that the effective epidemic threshold decreases with the average connectivity $\langle k \rangle$ under the assumption of homogeneous mixing~\cite{Anderson:1992}. Considering the heterogeneity of connectivity, the heterogeneous mean-field (HMF) theory ~\cite{Dorogovtsev:2008RMP,Pastor-Satorras:2001PRE,Barthelemy:2004PRL,Gomez2011} is employed to predict the effective epidemic threshold, which can be expressed as
\begin{equation}\label{SIR_HMF}
    \lambda_c^{HMF}=\frac{\langle k \rangle}{\langle k^2 \rangle - \langle k \rangle},
\end{equation}
where $\langle k \rangle$ and $\langle k^2 \rangle$ represent the first and second moments of degree distribution $P(k)$~\cite{Newman:Networks}, respectively. On networks with power-law scaling $P(k)\sim k^{-\gamma}$ where $\gamma$ represents the degree exponent~\cite{Newman:Networks,Albert:2002RMP}, the vanishing threshold for scale-free networks with $\gamma\leq3$ and the finite threshold for $\gamma>3$ are predicted by the HMF approach~\cite{Pastor-Satorras:2014}. The quenched mean-field (QMF) theory is proposed to attempt to improve the HMF theory, since the latter neglects the quenched structure of the network and dynamical correlations between the state of adjacent nodes~\cite{Givan:2011JTB}. In the QMF theory the actual quenched structure of the network is fully preserved, and the effective epidemic threshold is predicted as~\cite{Chakrabarti:2008ACM,Van Mieghem:2009ACM,Gomez:2010epl}
\begin{equation}\label{QMF}
    \lambda_c^{QMF}=\frac{1}{\Lambda_N},
\end{equation}
where $\Lambda_N$ represents the maximum eigenvalue of the adjacency matrix of a given network. However, the QMF result is even qualitatively not correct, because the vanishing threshold for power-law distributed networks with $\gamma > 3$ predicted by the QMF is in conflict with the visually numerical results~\cite{Castellano:2010PRL}.

It is worth noting that the results above have nothing to do with the recovery rate $\mu$, which determines the infection duration of a given disease. In fact, each disease has its own special infection duration. The symptoms of mumps resolve after 7 to 10 days~\cite{vaccine}. The time period between contracting the rabies disease and death can vary from less than one week to more than one year~\cite{Rabies2013}. Without treatment, the stage of HIV infection can last from about three years to over 20 years~\cite{HIV2006, HIV2001} (on average, about eight years). These diseases with different infection durations have lead to different levels of prevalence. According to the statistics, about 0.1 percent to 1 percent of the population are affected by mumps virus per year~\cite{vaccine}. Rabies causes about 26,000 to 55,000 deaths worldwide per year~\cite{Rabies2013}. Since its discovery, AIDS has caused an estimated dozens of million deaths worldwide~\cite{HIV2015}. Not only that, the duration of disease also affects the effective epidemic threshold. The existing studies~\cite{Pastor-Satorras:2014, Newman:2002PRE} generally think that if the infected nodes immediately recover after the infection process (i.e., $\mu=1$), the effective epidemic threshold coincides with the result of Eq.~(\ref{SIR_HMF}), and when the infected nodes can not immediately recover (i.e., $\mu<1$), the effective epidemic threshold is predicted by
\begin{equation}\label{SIR_Percolation}
\lambda_c=\frac{\langle k \rangle}{\langle k^2 \rangle - 2\langle k \rangle}.
\end{equation}

Although these studies have pointed out the difference in effective epidemic threshold for the cases with $\mu=1$ and $\mu<1$, the systematical studies on the effects of the recovery rate on the effective epidemic threshold is still insufficient. Moreover, the spreading dynamics is either asynchronous or synchronous updating process, which are two famous numerical methods for dynamics~\cite{WOLFRAM}. In different updating processes, how the recovery rate influences the spreading dynamics such as the effective epidemic threshold is long neglected. Here we develop an edge-based compartmental theory to derive the effective epidemic thresholds for the SIR model with arbitrary recovery rate, in both asynchronous and synchronous updating spreading processes~\cite{Miller:Math2011,Wang:PRE2014}. The proposed theory could be considered as supplementary to the existing theories, and it predicts that the effective epidemic threshold is independent of (decreases with) the recovery rate in asynchronous (synchronous) updating spreading processes. We further validate the theory based on extensive numerical simulations on synthetic and real-world networks. In most cases our theoretical predictions are in a good agreement with the numerical effective epidemic thresholds identified by the variability measure~\cite{crepey:2006PRE,Shu:2012}, which has been confirmed to be effective for identifying the SIR effective epidemic threshold~\cite{Shu:2015}. Although there exist some differences between the theoretical predictions and numerical results in networks with disassortative mixing, the theoretical effective epidemic threshold displays the same trend to that of the numerical effective epidemic threshold.

\section{Theory}\label{SEC:2}
To qualitatively understand the SIR dynamic with arbitrary recovery rate, we develop the edge-based compartmental theory based on Refs.~\cite{Miller:Math2011,Wang:PRE2014,Lidia}. On an uncorrelated and large sparse network, the SIR model can be described in terms of $S(t)$, $I(t)$ and $R(t)$, which represent the densities of the susceptible, infected, and recovered nodes at time $t$, respectively.

Let us now consider a randomly chosen node $u$ given that $u$ is in the cavity state initially, which means that it can not transmit any disease to its neighbors but can be infected by its neighbors. We define $\theta(t)$ to be the probability that a neighbor $v$ of $u$ has not transmitted the disease to $u$ along the edge connecting them up to time $t$. We assume that $\theta(t)$ is identical for all edges. Initially, a vanishingly small $\rho_0$ fraction of nodes are chosen to be infected and none of them transmits the disease to its neighbors, that is $\theta(0)=1$. According to the cavity theory~\cite{Karrer:PRE2010,Miller:Plos2013}, we obtain the probability that the node with degree $k$ is susceptible by time $t$ as $s(k,t)=\theta(t)^k$.
Averaging over all $k$, the density of susceptible nodes at time $t$ is given by
\begin{equation} \label{s_t}
S(t)=\sum_{k=0}^{\infty}P(k)\theta(t)^k.
\end{equation}
Obviously, to solve $S(t)$, we need to know $\theta(t)$. Since a neighbor of node $u$ may be susceptible, infected, or recovered, $\theta(t)$ can be expressed as
\begin{equation} \label{theta}
\theta(t)=\xi_S(t)+\xi_I(t)+\xi_R(t),
\end{equation}
where $\xi_S(t)$ [$\xi_I(t)$ or $\xi_R(t)$] is the probability that the
neighbor $v$ is in the susceptible (infected or recovery) state and has not
transmitted the disease to node $u$ through their connection.

According to the definition of cavity state above, the susceptible neighbor $v$ can only get the disease from its other neighbors when $u$ is in cavity state. Thus, node $v$ will keep susceptible at time $t$ with probability $\theta(t)^{k-1}$.
For uncorrelated networks, the probability that one edge from node $u$
connects with an node with degree $k^{\prime}$ is
$k^{\prime}P(k)/\langle k\rangle$. Summing over all possible $k^{\prime}$, we obtain
\begin{equation} \label{xi_S}
\xi_S(t)=\frac{\sum_{k^{\prime}}k^{\prime}P(k)
\theta(t)^{k^\prime-1}}{\langle k\rangle}.
\end{equation}

The time evolutions of $\xi_R$ are slightly different in the synchronous and asynchronous updating spreading processes. For the case of synchronous updating method, an infected node first may transmit the infection to its neighbors and then become recovered in a discrete time step. Since the infection and recovery events may happen consecutively, the notation $\xi_R$ means that the infected neighbor $v$ has not transmitted the disease to $u$ with probability $1-\beta$ via their connection, and simultaneously it recovers with probability $\mu$. Taking these into consideration, we get
\begin{equation} \label{xi_R}
\frac{d\xi_R(t)}{dt}=\mu(1-\beta)\xi_I(t).
\end{equation}
For the case of asynchronous updating method, the infection and recovery can not happen simultaneously, Eq.~(\ref{xi_R}) thus becomes
\begin{equation} \label{xi_R_A}
\frac{d\xi_R(t)}{dt}=\mu\xi_I(t).
\end{equation}
In the edge-based compartmental theory, the only difference between synchronous and asynchronous updating processes is the time evolution of $\xi_R$, as shown in Eqs.~(\ref{xi_R}) and (\ref{xi_R_A}). Therefore, we next introduce the theory based on the synchronous update method, unless explicitly stated.

Once the infected neighbor $v$ transmits the disease to $u$ successfully, $\theta(t)$ will change as
\begin{equation} \label{d_theta}
\frac{d\theta(t)}{dt}=-\beta\xi_I(t).
\end{equation}
For the case of synchronous updating process, combining Eqs.~(\ref{xi_R})-(\ref{d_theta}), and initial conditions $\theta(0)=1$ and $\xi_R(0)=0$, we obtain
\begin{equation} \label{xi_R_2}
\xi_R(t)=\frac{\mu[1-\theta(t)](1-\beta)}{\beta}.
\end{equation}
Substituting Eqs.~(\ref{xi_S}) and (\ref{xi_R_2}) into Eq.~(\ref{theta}), we get an expression for $\xi_I(t)$ in terms of $\theta(t)$, and then we
can rewrite Eq.~(\ref{d_theta}) as
\begin{equation} \label{d_theta_2}
\begin{split}
\frac{d\theta(t)}{dt}&=-\beta[\theta(t)-
\frac{\sum_{k^{\prime}} k^{\prime}P(k^{\prime})\theta(t)^{k^\prime-1}}
{\langle k\rangle}]\\
&+\mu[1-\theta(t)](1-\beta).
\end{split}
\end{equation}

We pay special attention to the final state of the epidemic spreading, where $\theta(t)$ will never change [i.e., $d\theta(t)/dt=0$] and thus we get
\begin{equation} \label{stady_theta}
\theta(\infty)=\frac{\sum_{k^{\prime}} k^{\prime}P(k^{\prime})
\theta(\infty)^{k^{\prime}-1}}{\langle k\rangle}+\frac{\mu[1-\theta(\infty)](1-\beta)}{\beta}.
\end{equation}
Substituting fixed point \textbf{$\theta^{*}(\infty)$} of Eq.~(\ref{stady_theta}) into Eq.~(\ref{s_t}), we can obtain the final susceptible density $S(\infty)$ and the final outbreak size $R(\infty)=1-S(\infty)$.

The value $\theta(\infty)=1$ is always a solution of Eq.~(\ref{stady_theta}). Define the right hand of Eq.~(\ref{stady_theta}) as $f(\theta(\infty))$. In order to get a nontrivial solution, the condition
\begin{equation} \label{threshold_condition}
\frac{d f(\theta(\infty))}{d (\theta(\infty))}|_{\theta(\infty)=1}>1
\end{equation}
must be fulfilled~\cite{Moreno:EPJB2002}. This relation implies that
\begin{equation}\label{threshold1}
\frac{\beta}{\mu}>\frac{\langle k \rangle}{\langle k^2\rangle+\mu\langle k\rangle -2\langle k\rangle}.
\end{equation}
This condition defines the effective epidemic threshold with synchronous updating
\begin{equation} \label{threshold}
\lambda_c^{sync}=\frac{\langle k\rangle}{\langle k^2\rangle -
2\langle k\rangle+\mu\langle k\rangle}.
\end{equation}
It can been seen from Eq.~(\ref{threshold}) that the effective epidemic threshold not only is affected by the network structure, but also decreases with the recovery rate $\mu$ in the synchronous updating process. Specially, Eq.~(\ref{threshold}) is exactly the HMF prediction when $\mu=1$, and it approaches Eq.~(\ref{SIR_Percolation}) when $\mu\rightarrow 0$.

In a similar way, we can solve out the effective epidemic threshold for the asynchronous updating method by substituting Eq.~(\ref{xi_R_A}) into the corresponding equations. Specifically, Eqs.~(\ref{xi_R_2}) and (\ref{stady_theta}) are rewritten as
\begin{equation} \label{xi_R_22}
\xi_R(t)=\frac{\mu[1-\theta(t)]}{\beta}
\end{equation}
and
\begin{equation} \label{stady_theta2}
\theta(\infty)=\frac{\sum_{k^{\prime}} k^{\prime}P(k^{\prime})
\theta(\infty)^{k^{\prime}-1}}{\langle k\rangle}+\frac{\mu[1-\theta(\infty)] }{\beta},
\end{equation}
respectively. Thus in the asynchronous updating process, the effective epidemic threshold of SIR model is given by
\begin{equation} \label{threshold2}
\lambda_c^{async}=\frac{\langle k\rangle}{\langle k^2\rangle
-2\langle k\rangle}.
\end{equation}
From Eq.~(\ref{threshold2}), we know that the effective epidemic threshold is only correlated with the topology of network, and irrelevant to the recovery rate.

\section{Main results}\label{SEC:3}
In the SIR model, a node of networks can be susceptible, infected, or recovered. In simulations, the SIR spreading processes are implemented by using both synchronous and asynchronous updating methods. The synchronous updating spreading process~\cite{Moreno:EPJB2002} is carried out as follows. At the beginning, one node is randomly selected as the initial infected (i.e., seed), and all other nodes are susceptible. In each time step $t$, each susceptible node $i$ becomes infected with probability $1-(1-\beta)^{n_i}$ if it has one or more infected neighbors, where $n_i$ is the number of its infected neighbors. In the same time, all infected nodes recover (or die) at rate $\mu$ and the recovered nodes acquire permanent immunity. Time increases by $\Delta t=1$, and the dynamical process terminates when there is no infected node in the network.

The asynchronous updating spreading process~\cite{WOLFRAM} is performed as follows. At time $t$, the number of infected nodes is denoted as $N_I(t)$, and the number of active edges (i.e., the edges connecting a susceptible node and an infected node) is recorded as $E_A(t)$. At each step, a randomly chosen infected node becomes recovered with probability $p_r=\mu N_I(t)/[\mu N_I(t)+\beta E_A(t)]$, otherwise, an active link is chosen at random and the susceptible node attached to it becomes infected with probability $1-p_r$. The time is updated as $t\rightarrow t+1/[\mu N_I(t)+\beta E_A(t)]$. The process terminates until there is no infected node in the network.

To numerically identify the effective epidemic threshold of the SIR model, we use the variability measure~\cite{crepey:2006PRE,Shu:2012}
\begin{equation}\label{variability}
\Delta=\frac{\sqrt{\langle \rho^2 \rangle - \langle \rho \rangle^2}}{\langle \rho \rangle}.
\end{equation}
The variability $\Delta$ exhibits a peak over a wide range of $\lambda$, and we estimate the numerical effective epidemic threshold $\lambda_c^{num}$ from the position of the peak of the variability. The validity of this numerical identification method for the SIR model has been confirmed in Ref.~\cite{Shu:2015}.

\subsection{Random regular networks}

We first consider the final outbreak size $R(\infty)$ as a function of $\lambda$ for different recovery rates on random regular networks (RRNs), where all nodes have exactly the same degree $k$. We investigate the effect of recovery rate on the SIR spreading dynamics by respectively using asynchronous and synchronous updating simulation methods.

\begin{figure}[h]
\begin{center}
\epsfig{file=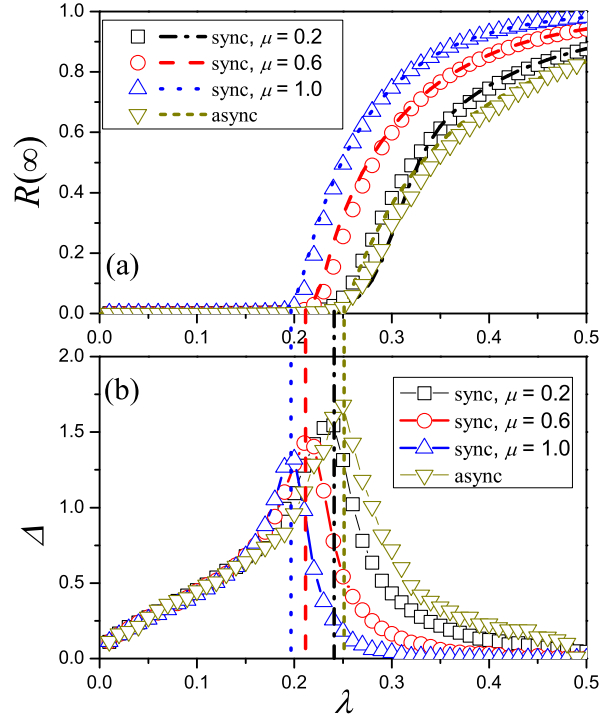,width=1\linewidth}
\caption{(Color online) Overview of SIR dynamics with different recovery rates $\mu$ on RRNs. (a) Final outbreak size $R$ vs. $\lambda$ for $\mu$=0.2 (squares and short dash line), 0.6 (circles and solid line) and 1.0 (up triangles and dot line) with synchronous updating, $\mu=0.2,0.6,1.0$ (down triangles and short dash line) with asynchronous updating, where symbols and lines represent the numerical results and theoretical predictions, respectively. (b) Variability $\Delta$ vs. $\lambda$ for $\mu$=0.2 (squares), 0.6 (circles) and 1.0 (up triangles) with synchronous updating, $\mu=0.2,0.6,1.0$ (down triangles) with asynchronous updating, respectively. The vertical lines point out the positions of numerical effective epidemic thresholds. The results are averaged over $10^2\times10^4$ independent realizations on $10^2$ different networks. The parameters are chosen as $N=10^4$ and $k=6$.}
\label{fig1}
\end{center}
\end{figure}

With asynchronous updating, Fig.~\ref{fig1} shows all simulation results for different values of recovery rate $\mu$ completely overlap with each other. Thus, we obtain a trivial conclusion: the effective epidemic threshold and final outbreak size is not affected by the recovery rate. According to the asynchronous updating method, the recovery probability $p_r$ and time interval $\Delta t$ can be rewritten as $N_I(t)/[N_I(t)+\lambda E_A(t)]$ and $1/\mu[N_I(t)+\lambda E_A(t)]$. When the effective transmission rate $\lambda=\beta/\mu$ is fixed, the change of recovery rate $\mu$ does not affect the recovery probability $p_r$ and infection probability $1-p_r$, while only alters the relative size of time scale $\Delta t=1/\mu[N_I(t)+\lambda E_A(t)]$. Therefore, the recovery rate does not change the effective epidemic threshold and final outbreak size in the asynchronous updating spreading process. The developed theory can describe the phenomena very well.

With synchronous updating, Fig.~\ref{fig1} (a) shows that the final outbreak size for small recovery rate (e.g., $\mu=0.2$) is obviously smaller than that for large recovery rate (e.g., $\mu=1.0$) at the same effective transmission rate $\lambda$, and the simulated results can agree fairly well with the theoretical predictions from the edge-based  compartmental theory. This phenomenon indicates that the recovery rate will have a remarkable effect on the SIR epidemic dynamics with synchronous updating. In Fig.~\ref{fig1} (b), we further plot the variability $\Delta$ as a function of $\lambda$ to numerically identify the effective epidemic threshold for the synchronous updating method. The results show that the peak of the $\Delta$ gradually shifts to the left as the recovery rate $\mu$ increases. In other words, the effective epidemic threshold increases with the decrease of $\mu$ when the synchronous updating method is used.

From Fig.~\ref{fig1}, we know the recovery rate only alters the time scale of asynchronous updating spreading dynamics, but significantly affects the synchronous updating spreading dynamics. Moreover, we find that the synchronous updating spreading breaks more easily and has a greater final outbreak size compared with the case of asynchronous updating (give a qualitative explanation later). Next, we only focus on the effect of recovery rate on the synchronous updating spreading dynamics, unless explicitly stated.

\begin{figure}[t]
\begin{center}
\epsfig{file=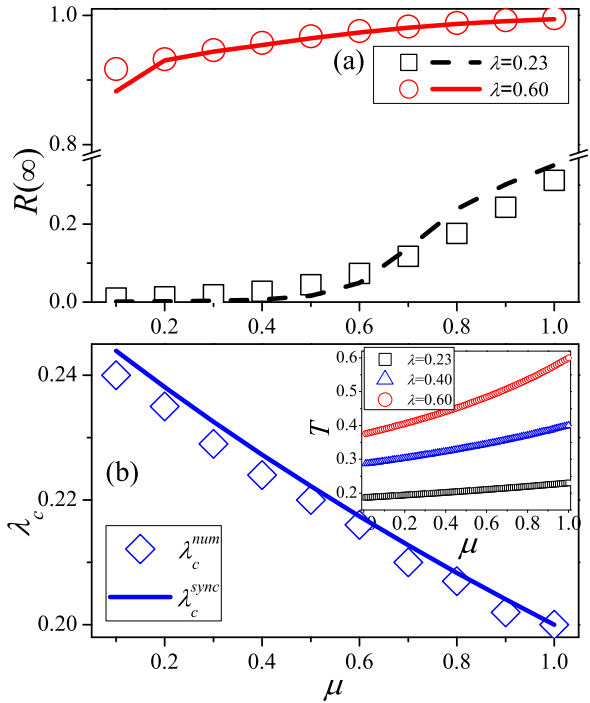,width=1\linewidth}
\caption{(Color online) In the spreading dynamics with synchronous updating, final outbreak size $R$ and effective epidemic threshold $\lambda_c$ as a function of recovery rate $\mu$ on RRNs. (a) $R$ vs. $\mu$ for $\lambda=0.23$ (squares and dash line) and $\lambda=0.60$ (circles and solid line), respectively. (b) $\lambda_c$ vs. $\mu$. The inset of (b) shows the mean transmission probability $T$ (through one edge of an infected node before it recovers) versus $\mu$ at $\lambda=0.23$ (black squares), $\lambda=0.40$ (blue up triangles) and $\lambda=0.60$ (red circles). In each figure, symbols and lines represent the numerical and theoretical results, respectively. The parameters are chosen as $N=10^4$ and $k=6$. The results are averaged over $10^2\times10^4$ independent realizations on $10^2$ different networks.}
\label{fig2}
\end{center}
\end{figure}

Given the value of $\lambda$, we show the final outbreak size as a function of $\mu$ under the synchronous updating method in Fig.~\ref{fig2}, where a small $\lambda=0.23$ and a large $\lambda=0.60$ are considered, respectively. As shown in Fig.~\ref{fig2} (a), for the small value of $\lambda$, the final outbreak size is very tiny when $\mu$ is small, while the epidemic can infect a finite proportion of nodes for large $\mu$. For the large value of $\lambda$, the final outbreak size for the small $\mu$ is still smaller than that for large $\mu$.

In Fig.~\ref{fig2} (b), one see that the effective epidemic threshold decreases with $\mu$ both numerically and theoretically. The consistency of the simulated results and theoretical predictions confirms the validity of the edge-based compartmental theory. To qualitatively understand these phenomena, the inset of Fig.~\ref{fig2} (b) shows the mean transmission probability $T$ through one edge of an infected node before it recovers, where $T=\sum_{t=1}^\infty [(1-\mu)(1-\beta)]^{t-1}\beta$. Once the value of $\lambda$ is given, $T$ increases with the recovery rate $\mu$, which is greater than $T=\lambda/(1+\lambda)$ for the case of asynchronous updating~\cite{Pastor-Satorras:2014}. In other words, the mean infection ability of a single infected node is enlarged by large recovery rate. This effect leads to the decrease of the effective epidemic threshold with the recovery probability $\mu$. It means that when the recovery probability is large, the epidemic can outbreak even if $\lambda$ is small, while when the recovery probability is small, the epidemic can outbreak just for a large $\lambda$.

\subsection{Scale-free networks}
\begin{figure}
\begin{center}
\epsfig{file=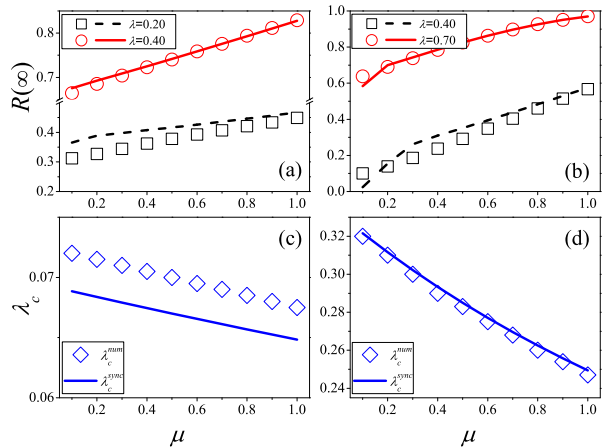,width=1.1\linewidth}
\caption{(Color online) In the spreading dynamics with synchronous updating, final outbreak size $R$ and effective epidemic threshold $\lambda_c$ as a function of recovery rate $\mu$ on SFNs with degree exponents $\gamma=2.5$ [(a) and (c)] and $\gamma=4.0$ [(b) and (d)], where the network size is set as $N=10^4$. (a) $R$ vs. $\mu$ for $\lambda=0.2$ (squares and dash line) and $\lambda=0.4$ (circles and solid line), respectively. (b) $R$ vs. $\mu$ for $\lambda=0.4$ (squares and dash line) and $\lambda=0.7$ (circles and solid line), respectively. (c)-(d) $\lambda_c$ vs. $\mu$. In each figure, symbols and lines respectively represent the numerical and theoretical results. We perform $10^2\times 10^4$ independent realizations on $10^2$ different networks.}
\label{fig3}
\end{center}
\end{figure}

We further consider the SIR epidemic dynamics with arbitrary recovery rate on scale-free networks with power-law degree distribution $P(k)\sim k^{-\gamma}$, where the synchronous updating method is implemented. We build scale-free networks (SFNs) based on the configuration model~\cite{Newman:Networks}. The so-called structural cutoff~\cite{Boguna:2004EPJB} $k_{max}\sim N^{1/2}$ is considered to constrain the maximum possible degree $k_{max}$ on SFNs, where the degree-degree correlations vanish in the thermodynamic limit. Fig.~\ref{fig3} shows the final outbreak size and effective epidemic threshold as a function of $\mu$ for SFNs with $\gamma=2.5$ and $\gamma=4.0$. Like the case on RRNs, the final outbreak size increases with recovery rate when the value of $\lambda$ is given [see Figs.~\ref{fig3} (a) and (b)], and the effective epidemic threshold decreases with $\mu$ [see Figs.~\ref{fig3} (c) and (d)]. Due to the weak degree heterogeneity of SFNs with $\gamma=4.0$, the effective epidemic threshold decreases more rapidly with $\mu$ in such networks than that for SFNs with $\gamma=2.5$. The theoretical predictions are very close to the simulated results for SFNs with $\gamma=4.0$, while there are some difference between them for SFNs with $\gamma=2.5$, because the disassortative degree-degree correlations still exist in such networks with finite size~\cite{Boguna:2004EPJB}.

\subsection{Real-world networks}
\begin{figure}[t]
\begin{center}
\epsfig{file=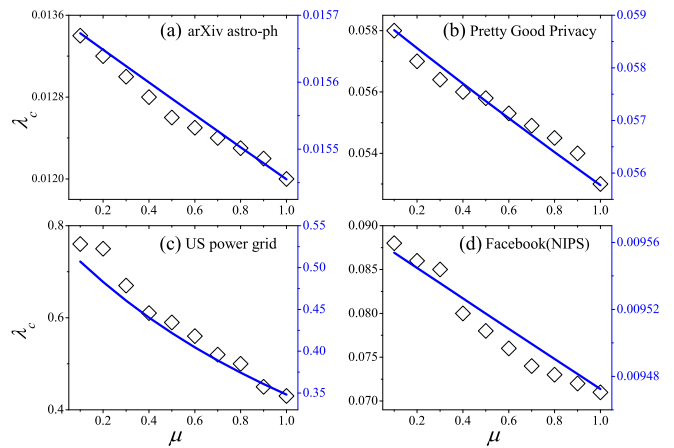,width=1\linewidth}
\caption{(Color online) In the spreading dynamics with synchronous updating, the effective epidemic threshold $\lambda_c$ as a function of recovery rate $\mu$ on real-world networks. (a) arXiv astro-ph network. (b) Pretty Good Privacy network. (c) US power grid network. (d) Facebook (NIPs) network. In each figure, the symbol represents the numerical effective epidemic threshold, whose value is shown by the left scale mark, and the solid line represents the theoretical effective epidemic threshold, whose value is shown by the right scale mark. We perform $10^6$ independent realizations on each real network.}
\label{fig4}
\end{center}
\end{figure}

To further study the cases of real-world networks with synchronous updating, we consider four typical real networks, which are arXiv astro-ph~\cite{astroph}, Facebook (NIPS)~\cite{Facebook}, Pretty Good Privacy~\cite{Boguna:PRE2004}, and US power grid~\cite{Watts:Nature1998}. Several structural characteristics of this four real example networks are presented in Table~\ref{table}, where the difference among these networks implies the complexity of real network structure to a certain extent. The numerical and theoretical thresholds of these networks are shown in Fig.~\ref{fig4}, where the results again show that both the theoretical and numerical thresholds decrease with the recovery rate. Although the theoretical predictions agree relatively well with the numerical thresholds for assortative networks, there is an obvious gap between them for the Facebook (NIPs) network showing disassortative mixing. Compared with the cases on other real networks, we can find that the effective epidemic threshold changes more rapidly with the recovery rate on the US power grid network. The difference in the variation of the effective epidemic threshold for different networks could be attributed to the complexity of real network structures.\\

\begin{table}
\scriptsize
\caption{Structural characteristics of four real-world networks. $N$ is the network size, $k_{max}$ is the maximum degree, $\langle k\rangle$ is the average degree, $c$ is the clustering coefficient, $r$ is the Pearson
correlation coefficient, and $d$ stands for the diameter of the network.
}\label{table}
  \begin{tabular}{|l|c|c|c|c|c|c|c|c|c|c|c|}
    \hline
    \hline
    Network & Category & $N$ & $k_{max}$ & $\langle k\rangle$ & $c$ & $r$ & $d$\\
    \hline
    arXiv astro-ph~\cite{astroph} & Coauthorship & 17903 & 504 & 22.004 & 0.633 & 0.201 & 14\\
    Pretty Good Privacy~\cite{Boguna:PRE2004} & OnlineContact & 10680 & 206 & 4.558 & 0.266 & 0.239 & 24\\
    US power grid~\cite{Watts:Nature1998} & Infrastructure & 4941 & 19 & 2.669 & 0.080 & 0.003 & 46\\
    Facebook(NIPS)~\cite{Facebook} & Social & 2888 & 769 & 2.064 & 0.027 & -0.668 & 9\\
    \hline
    \hline
  \end{tabular}
\end{table}

\section{Conclusion and discussion}\label{SEC:4}
In this paper, we have made a detailed study of the SIR model with arbitrary recovery rate. Firstly, we developed an edge-based compartmental theory to predict the final outbreak size and effective epidemic threshold for arbitrary recovery rate. Two basic updating methods are considered: asynchronous updating and synchronous updating. For the case of asynchronous updating, the recovery rate only alters the time scale of SIR spreading dynamics, but does not affect the phase transition and final state. However, for the case of synchronous updating, the developed theory predicted that the effective epidemic threshold decreases with the recovery rate, and the final outbreak size for small recovery rate is obviously smaller than that for large recovery rate when the value of $\lambda$ is given.

To verify the theoretical predictions, we considered the SIR dynamics on RRNs with constant degree. With asynchronous updating, both the effective epidemic threshold an final outbreak size remain unchanged for different recovery rates, while the obvious difference in final outbreak size for different values of $\mu$ is observed in the synchronous updating spreading process. We numerically identified the effective epidemic threshold $\lambda_c$ with the variability measure, which has been confirmed to be effective for identifying the SIR effective epidemic threshold, and found that $\lambda_c$ indeed decreases with $\mu$ for the case of synchronous updating. As the infection and recovery events may happen consecutively in the synchronous updating process, the mean infection ability of a single infected node is enlarged by large recovery rate. The results showed good agreements between theoretical predictions and simulated results on RRNs. To explore the university of these conclusions, we further carry on these studies on scale-free and real-world networks, where the similar phenomena were observed. Although a certain gap between the theoretical predictions and numerical thresholds still exists for some networks with disassortative mixing patterns, the developed theory can indeed give a relatively accurate prediction of the effective epidemic threshold in most cases.

We have theoretically and numerically demonstrated that with synchronous updating, the effective epidemic threshold and final outbreak size of the SIR dynamical processes are affected by the recovery rate. The results showed that if one ignores the effect of the recovery rate, it may leads to the misunderstanding of SIR synchronous updating dynamics with effective spreading rate $\lambda=\beta/\mu$. For example, the existing studies generally considered the effective SIR epidemic threshold with constant value $\mu=1$, while the effective epidemic threshold decreases with $\mu$ actually. Our work supplemented the existing studies on the effective epidemic threshold, and provided us with deeper understanding on the phase transition of epidemic dynamics. It should be noted that the SIR epidemic of synchronous updating outbreaks more easily than asynchronous updating, only when the recovery rate is close to zero, the final state of synchronous updating tends to that of asynchronous updating. Moreover, there still exists a certain gap between the theoretical predictions and simulated results for some disassortative networks, and thus more accurate analytic approximation of the effective epidemic threshold (e.g., message-passing approach~\cite{Shrestha:PRE2015,Lokhov:PRE2015}) for SIR dynamics with arbitrary recovery rate remains an important problem.
\\
\\
\begin{acknowledgments}
This work was partially supported by National Natural Science Foundation of China (Grant Nos.~11105025, 11575041, 61433014, 61501358).
\end{acknowledgments}

\end{document}